\documentclass{emulateapj}
\usepackage{apjfonts}

\usepackage{amssymb}
\usepackage{natbib}
\usepackage{graphicx}

\slugcomment{}
\shorttitle{Cen X-3 over two binary orbits}
\shortauthors{Suchy et al.}

\begin{document}

\title{Pulse phase resolved analysis of the High Mass X-Ray Binary
  Centaurus~X-3 over two binary orbits} 

\author{Slawomir~Suchy\altaffilmark{1},
Katja~Pottschmidt\altaffilmark{1,2,3},
J\"orn~Wilms\altaffilmark{4},
Ingo~Kreykenbohm\altaffilmark{5,6},
Gabriele~Sch\"onherr\altaffilmark{7},
Peter~Kretschmar\altaffilmark{8},
Vanessa~McBride\altaffilmark{9},
Isabel~Caballero\altaffilmark{5},
Richard~E.~Rothschild\altaffilmark{1},
Victoria~Grinberg\altaffilmark{10}}

\altaffiltext{1}{University of California, San Diego, Center for
  Astrophysics and Space Sciences, 9500 Gilman Dr., La Jolla, CA
  92093-0424, USA, \{ssuchy,kpottschmidt,rrothschild\}@ucsd.edu}
 \altaffiltext{2}{Department of Physics, University of
  Maryland Baltimore County, 1000 Hilltop Circle, Baltimore, MD 21250,
  USA} 
\altaffiltext{3}{NASA Goddard Space Flight Center,
  Astrophysics Science Division, Code 661, Greenbelt, D 20771, USA,
  katja@milkyway.gsfc.nasa.gov} 
\altaffiltext{4}{Dr.\
  Karl-Remeis-Sternwarte, Astronomisches Institut, Sternwartstr.\ 7,
  96049 Bamberg, Germany, joern.wilms@sternwarte.uni-erlangen.de}
  \altaffiltext{5}{Institut f\"ur Astronomie und Astrophysik, Abt.\
  Astronomie, Sand 1, 72076 T\"ubingen, Germany,
  \{kreyken,isabel\}@astro.uni-tuebingen.de}
  \altaffiltext{6}{\textsl{INTEGRAL} Science Data Centre, Chemin
  d'\'Ecogia 16, 1290 Versoix, Switzerland,
  ingo.kreykenbohm@obs.unige.ch} \altaffiltext{7}{Astrophysikalisches
  Institut Potsdam, An der Sternwarte 16, 14482 Potsdam, Germany,
  gschoenherr@aip.de} \altaffiltext{8}{European Space Agency, European
  Space Astronomy Centre, P.O. Box 78, 28691 Villanueva de la
  Ca$\tilde{\mathrm{n}}$ada, Madrid, Spain, peter.kretschmar@esa.int}
  \altaffiltext{9}{School of Physics and Astronomy, Southampton
  University, Southampton, SO17 1BJ, United Kingdom,
  vanessa@astro.soton.ac.uk} \altaffiltext{10}{Fakult\"at f\"ur
  Physik, Ludwig-Maximilians-Universit\"at M\"unchen, Schellingstr.\
  4, 80799 M\"unchen, Germany, victoria.grinberg@physik.lmu.de}
  \email{ssuchy@ucsd.edu}

\begin{abstract}
  We present a detailed analysis of observations of the high mass
  X-ray binary Cen~X-3 spanning two consecutive binary orbits
  performed with the \textsl{RXTE} satellite in early March 1997.
  During this time Cen~X-3 had a luminosity of
  $L_{2-10\,{\mathrm{keV}}} \sim 4$--$5 \times 10^{37}
  {\mathrm{erg\,s}}^{-1}$ and a pulse period of $4.814$\,s. The PCA
  and HEXTE light curves both show a clear reduction in count rate
  after mid-orbit for both binary revolutions. We therefore analyze
  two broad band spectra for each orbit, before and after
  mid-orbit. Consistent with earlier observations these four joint PCA
  and HEXTE spectra can be well described using a phenomenological
  pulsar continuum model, including an iron emission line and a
  cyclotron resonance scattering feature.  While no strong spectral
  variations were detected, the second half of orbit 2 shows a
  tendency toward being softer and more strongly absorbed. In order to
  follow the orbital phase-dependent evolution of the spectrum in
  greater detail, we model spectra for shorter exposures, confirming
  that most spectral parameters show either a gradual or sudden change
  for the second half of the second orbit. A comparison with a simple
  wind model indicates the existence of an accretion wake in this
  system. We also present and discuss high resolution pulse profiles
  for several different energy bands, as well as their hardness
  ratios. PCA and HEXTE spectra were created for 24 phase bins and
  fitted using the same model as in the phase averaged
  case. Systematic pulse phase-dependent variations of several
  continuum and cyclotron line parameters were detected, most notably
  a significant increase of the cyclotron line energy during the early
  rise of the main peak, followed by a gradual decrease. We show that
  applying a simple dipole model for the magnetic field is not
  sufficient to describe our data.
\end{abstract}


\section{Introduction} \label{sec:intro} The pulsar nature of the
accreting high mass X-ray binary (HMXB) Centaurus~X-3 was first
discovered by \textsl{UHURU} \citep{schreier72,giacconi71}, measuring
a spin period of $\sim$ 4.8\,s and an orbital period of $\sim$
2.1\,days. The binary system consists of a neutron star with a mass of
$1.21\pm 0.21\,{\mathrm{M}}_\odot$ accompanied by an O\,6--8~III
supergiant star with a mass of $20.5\pm0.7 {\mathrm{\,M}}_\odot$
\citep{hutchings79,ash99}. The distance of the binary system has been
estimated to be roughly 8\,kpc with a lower limit of 6.2\,kpc
\citep{krzeminski74} and an eccentricity of $\leq 0.0016$
\citep{bildsten97}. \citet{day91} used dust scattering measurements to
derive a distance of $5.4\pm0.3$\,kpc, a result only marginally
consistent with \citet{krzeminski74} within given uncertainties.  In
order to allow for easier comparison with earlier publications we will
adopt the widely used distance of 8\,kpc. HMXBs generally show a
strong stellar wind which can supply a continuous tidal stream to the
compact object when the companion star is close to filling its Roche
lobe \citep{petterson78, stevens88, blondin91}. \citet{day93}
discussed that in a system like Cen~X-3 an X-ray excited wind would be
sufficient as a possible mechanism for mass transfer able to sustain a
constant mass flux.  A system with a combined mass transfer
mechanism is, e.g., GX~301$-$2 \citep{leahy02, leahy07}, where the folded
\textsl{RXTE}/ASM light curve was best modeled by a stellar wind with 
an additional gas stream as described by \citet{stevens88}.
In addition evidence for an accretion disk in Cen~X-3 can be
found: \citet{tjemkes86} compared the observed light curve of Cen~X-3
with modeled binary light curves and achieved best results when
including an accretion disk. The observed overall spin-up trend of
1.135\,ms/yr with fluctuations on timescales of a few years
\citep{tsunemi96} also indicates the presence of an accretion disk,
attributed to Roche lobe overflow. A realistic picture of the mass
transfer mechanism in Cen~X-3 would therefore be a combination of
disk accretion with an excited stellar wind, as proposed by
\citet{petterson78}. 
 
An absorption feature in the X-ray spectrum at $\sim$30\,keV,
interpreted as a cyclotron resonance scattering feature (CRSF) also
known as cyclotron line, was first observed in Cen~X-3 with
\textsl{Ginga} by \citet{nagase92}. \textsl{BeppoSAX} and
\textsl{RXTE}/HEXTE confirmed this feature at 28\,keV and 30\,keV,
respectively \citep{santa98,heindl99}. It is assumed to be the
fundamental line and so far no higher line harmonics have been
observed for this source.  CRSFs are generated due to the interaction
of photons with electrons which are quantized in Landau levels, the
discrete energy states populated by electrons in strong magnetic
fields. Photons of energies corresponding to the separation of the
Landau levels are absorbed and re-emitted by the electrons. The net
effect is that photons from the accretion column above the magnetic
poles that have energies near the CRSF centroid energy are scattered
out of the line of sight and/or redistributed in energy.  These
features can be used to measure directly the magnetic field strength
in the line-forming region since the fundamental cyclotron line energy
is given by
\begin{equation} 
E_{\mathrm{cyc}} = \frac{1}{1+z}\frac{\hbar e B}{m_{\mathrm{e}}} =
\frac{11.6 {\mathrm{keV}}}{1+z} B_{12},
\end{equation}
where $B_{12}$ is the magnetic field strength near the neutron star
surface, in units of $10^{12}$\,Gauss and $z$ is the gravitational
redshift \citep{canuto77}. Assuming a typical neutron star mass of
$1.4 {\mathrm{\,M}}_\odot$ and neutron star radius of 10\,km gives $z
=0.3$. Cen~X-3's cyclotron line at $\sim$ 30\,keV then corresponds to
a magnetic field of $\sim 3.4\times 10^{12}$\,G.

\cite{kohmura01} analyzed the PCA data of this observation in the
energy band $2-13$\,keV which has a high time resolution of
15.625\,ms. They saw a delay in the time variability of the iron K
band by $0.39\pm 0.10$\,ms, and concluded that these photons were
reprocessed away from the neutron star and are probably part of the
matter accreting onto the NS.  A preliminary analysis by
\citet{heindl99} of the HEXTE data presented in this work concentrated
on basic pulse phase resolved spectroscopy, dividing the pulse profile
into four broad phase bins, and provided a first indication of a
possible increase of the cyclotron line energy during the rise of the
main pulse. Systematic variations of the CRSF centroid energy over
pulse phase have also been observed in other sources, e.g., Her X-1
\citep{soong90,gruber01}, Vela~X-1 \citep{labarbera03,kreykenbohm02},
4U\,0352$+$309 \citep{coburn01}, 4U\,1538$-$52 \citep{clark90}, and
GX~301$-$2 \citep{kreykenbohm04}.  The origin of these changes is not
yet well understood.  Straightforward mechanisms like the angular
dependence of relativistic corrections to the cyclotron energy may not
be sufficient to explain the observed variations.  Higher harmonics
rarely, if ever, are integer multiples of the fundamental line
energy. In some sources this can be explained by invoking the proper
non-integer line spacing expected from the relativistic cross section
\cite[e.g.,][]{pottschmidt05}. However, more complicated mechanisms
such as a B-field gradient at the magnetic pole, have also been
invoked \citep{nishimura05}. In addition a correlation between CRSF
energy and luminosity can be observed in several sources
\citep[e.g.,][]{mihara95,mowlavi06,labarbera05,staubert07,
nakajima06}.

In the following we further characterize the phase dependence of the line
parameters in Cen~X-3 using the exceptionally long -- spanning
$\sim$4\,days, i.e., two binary revolutions -- \textsl{RXTE}
observation of the source in 1997.  \citet{burderi00}, in the
following referred to as B00, observed Cen~X-3 simultaneously with
\textsl{BeppoSAX} overlapping 11\,hours of \textsl{RXTE} observations
during the first binary orbit. We will compare our results with B00
whenever possible.

We use combined PCA and HEXTE data for the analysis which allows us to
constrain the continuum shape of the spectrum and the CRSF parameters
much more strongly than with HEXTE alone. We compare our analysis with
other cyclotron line sources and apply a newly developed physical
model for the CRSF to the data. For the analysis over the orbit we
compare with a simplified wind model to discuss the evolution of
$N_{\mathrm{H}}$ through the observation.

The remainder of this paper is organized as follows: basic observation
parameters as well as the data reduction are described in
\S\ref{sec:data}. In \S\ref{sec:light} we describe the PCA and HEXTE
light curves observed over the two binary orbits.  Phase averaged
spectra for the first and second halves of each orbit and their best
fit models are presented in \S\ref{sec:phase_avg}, as well as the
temporal evolution of the spectral parameters with a resolution of
roughly one \textsl{RXTE} orbit throughout the two Cen X-3 binary
orbits.  \S\ref{sec:phase_res} is dedicated to the pulse phase
resolved analysis, including the presentation of pulse profiles in
multiple energy bands and modeling of phase resolved spectra. The
results are discussed in \S\ref{sec:discussion}.

\section{Observation and data reduction}\label{sec:data}

\textsl{RXTE} observed Cen~X-3 with the Proportional Counter Array
\citep[PCA;][]{jahoda05} and the High Energy X-ray Timing Experiment
\citep[HEXTE;][]{rothschild98} between 1997 February 28 and March
3. The observation, proposal ID P20104, covered two consecutive binary
orbits between eclipses, excluding the time of eclipse between
orbits. The total exposure time of 310\,ks was split into 12 shorter
data sets (``ObsIds'') of $\sim$ 8 hours each.

The PCA consists of five Proportional Counter Units (PCUs) covering a
nominal energy range from 2 to 60\,keV. For $\sim$ 90\% of the time
all PCUs were collecting data during the observation, with PCU~3 being
turned off during the remaining time. The PCA background was modeled
using the ``Sky-VLE'' model. The HEXTE instrument consists of two
independent clusters, each containing four NaI(Tl)/CsI(Na) phoswich
scintillation counters, covering an energy range from 15 to
250\,keV. Each cluster pointed alternately at the source and at the
background, switching every 16\,s. At any given time one of the
clusters was on-source.

The data were reduced using HEASOFT version 6.0.6., applying standard
extraction criteria: a pointing offset of $< 0.01^{\circ}$ from the
nominal source position, an exclusion time for the South Atlantic
Anomaly (SAA) of 30\,min, and a source elevation of $> 10^\circ$. We
imposed a maximum ``electron ratio'' of 0.15 in order to filter for
background flares, taking into account a slightly elevated electron
ratio level due to the influence of the bright target source.
Depending on the analysis method, further grouping and filtering of
the resulting good time intervals (GTIs) were applied in order to
derive the PCA and HEXTE light curves and spectra. Details are
described below, including the respective choices of data modes and
binning options. All HEXTE data products were deadtime corrected using
the HEASOFT ftool \texttt{hxtdead}. Since precise absolute flux
measurements are beyond the scope of this paper, correction for the
PCA deadtime of a few percent has not been carried out. Systematic
uncertainties of 0.5\% were included in the PCA spectral data. All
high time resolution products were converted into the frame of
reference at barycenter. The orbital ephemeris presented by
\citet{bildsten97} was adopted for the binary star correction. For
spectral modeling we used XSPEC version 11.3.2p \citep{arnaud96}.

\begin{figure*}
\begin{center}
\includegraphics[width=0.8\textwidth]{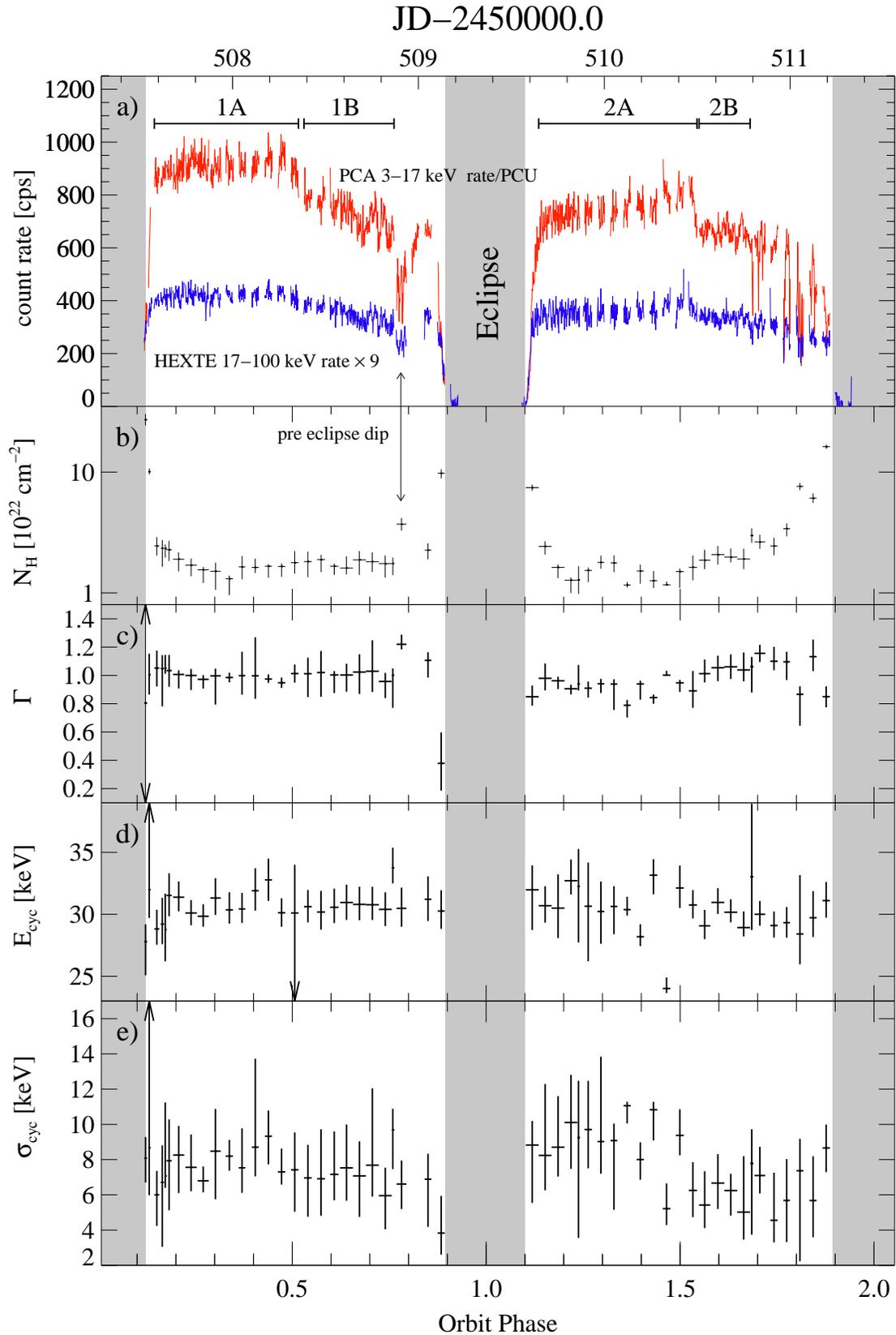}
\caption{(a) Background subtracted 3--17\,keV PCA (top) and
17--100\,keV HEXTE (bottom) light curves (bintime 128\,s). Shaded
region indicates the eclipse of the compact object.  Error bars are
not plotted but are smaller than any variations visible. Horizontal
lines above the data indicate the time intervals used for creating the
time averaged spectra analyzed in \S\ref{sec:timeavg}. (b)--(e)
Evolution of selected spectral parameters over two binary orbits. See
text for model details. While the spectrum is stable -- away from
enhanced (pre-)eclipse $N_{\mathrm{H}}$ -- over the first orbit, the
behavior changes significantly after the mid-orbit break of the second
orbit. Data points from spectral analysis during eclipse are ignored.}
\label{fig:light}
\end{center}
\end{figure*}

\section{Broad band orbital light curve}\label{sec:light}

We extracted PCA and HEXTE light curves for all available ObsIDs. For
the PCA, ``standard2f'' mode data were used, selecting the well
calibrated energy range from 3--17\,keV and the highest possible time
resolution for this data mode of 16\,s. Corresponding background model
light curves were created. In the case of HEXTE we use science event
mode data from 17--100\,keV, with the source not being significantly
detected at higher energies. Since the HEXTE source and background
measurements for each cluster are performed quasi-simultaneously due
to the 16\,s rocking cycle, the high time resolution background for
HEXTE light curves with less than 16\,s time resolution is based on
interpolating the measured background rates. For each 1\,s
time-resolved data bin in a 16\,s segment we performed a linear
interpolation of the background, using the neighboring 16\,s
background pointings\footnote{More precisely the background is
measured during 12\,s, preceded and followed by 2\,s during which the
cluster is moving between on- and off-source
\citep{rothschild98}.}. As HEXTEs cluster B only has 3 fully
functional detectors, the count rate for this cluster was normalized
to cluster A.

Fig.~\ref{fig:light}\emph{a} shows the background subtracted PCA and
HEXTE light curves (both rebinned to a resolution of 128\,s for
display purposes).  The PCA light curve has been normalized to one PCU
and the HEXTE light curve consists of the summed contributions of both
clusters and has been multiplied by a factor of nine, for display
purposes. The same general features are visible above and below
17\,keV, indicating the absence of extreme spectral variability over
the binary orbit. For orbital phase $\phi_{\mathrm{orb}}$ $\gtrsim$~0.5
both orbits show a clear drop in the count rates. In addition the
source was fainter overall during the second orbit, with the brighter
first half of orbit~2 approximately at the same level as the less
bright second half of orbit~1. The eclipse covers $\sim$ 20\% of the
orbital period. Pre-eclipse dips can be observed in both orbits.
\citet{takeshima91} and \citet{nagase92} also observed a drop in
luminosity for the second part of the orbit, calling it a pre-eclipse
dip. In our case we distinguish between before and after mid-orbit
break and use the term pre-eclipse dip for relative short count rate
drops, slightly longer than one satellite orbit.

In the first orbit one single pre-eclipse dip is visible around
$\phi_{\mathrm{orb}} = 0.75$ (JD 2450508.4) where the PCA rate drops by a
factor of two. Unfortunately, no HEXTE data are available in the
archive for the egress of this pre-eclipse because the instrument was
not turned on (PCA data are available). In the second orbit dipping
activity starts around the same $\phi_{\mathrm{orb}}$ (JD 2450510.8),
however, all pre-eclipse dips are partially obscured by data gaps due
to SAA passages and earth occultations, so that no detailed analysis
is possible.

\begin{deluxetable}{lcccc}
\tablecolumns{5} \tablecaption{Best fit parameters for phase averaged
spectra over first and second halves of each binary orbit.}\tablehead{
\colhead{} & \colhead{1A} & \colhead{1B} & \colhead{2A} &
\colhead{2B}} 
\startdata $N_{\mathrm{H}}$ [$10^{22}$cm$^{-2}$] &
$1.6^{+0.4}_{-0.2}$ & $1.8^{+0.5}_{-0.5}$ & $1.3^{+0.4}_{-0.2}$ &
$2.4^{+0.5}_{-0.4}$ \\ 
$E_{\mathrm{cut}}$ [keV] & $11.1^{+1.5}_{-2.3}$
& $4.8^{+9.7}_{-4.8}$ & $7.7^{+4.0}_{-2.5}$ & $20.9^{+6.6}_{-7.1}$ \\
$E_{\mathrm{fold}}$ [keV] & $7.2^{+0.2}_{-0.3}$ & $7.7^{+0.8}_{-0.5}$
& $7.2^{+0.1}_{-0.4}$ & $6.6^{+1.0}_{-2.3}$ \\ 
$E_{\mathrm{cyc}}$
[keV] & $30.7^{+0.5}_{-0.4}$ & $31.2^{+0.7}_{-0.7}$ &
$31.5^{+0.6}_{-0.5}$ & $30.3^{+0.7}_{-0.6}$ \\ 
$\sigma_{\mathrm{cyc}}$
[keV] &$6.4^{+1.0}_{-0.8}$ & $4.3^{+2.8}_{-1.6}$ & $6.3^{+0.9}_{-0.5}$
& $7.3^{+1.4}_{-1.6}$ \\ 
$\tau_{\mathrm{cyc}}$ &
$0.67^{+0.17}_{-0.07}$ & $0.48^{+0.27}_{-0.09}$ &
$0.61^{+0.09}_{-0.06}$ & $0.97^{+1.52}_{-0.34}$ \\
 $\Gamma$ &
$0.92^{+0.06}_{-0.04}$ & $0.88^{+0.20}_{-0.15}$ &
$0.76^{+0.05}_{-0.06}$ & $1.16^{+0.17}_{-0.09}$ \\ 
$A_{\mathrm{pl}} \dagger$ %[ph ${\mathrm{cm}}^{-2}\,{\mathrm{s}}^{-1}$] 
& $0.70^{+0.07}_{-0.05}$ &
$0.75^{+0.33}_{-0.18}$ & $0.52^{+0.12}_{-0.08}$&
$0.58^{+0.10}_{-0.06}$ \\ 
$E_{\mathrm{Fe}}$ [keV] &
$6.60^{+0.06}_{-0.05}$ & $6.57^{+0.05}_{-0.06}$ &
$6.61^{+0.06}_{-0.05}$ & $6.55^{+0.06}_{-0.06}$ \\
$\sigma_{\mathrm{Fe}}$ [keV] & $0.13^{+0.18}_{-0.13}$ &
$0.22^{+0.17}_{-0.22}$ & $0.09^{+0.21}_{-0.09}$ &
$0.27^{+0.15}_{-0.26}$ \\ 
$A_{\mathrm{Fe}} \dagger$ %[10^{-2}$ ph${\mathrm{cm}}^{-2}\,{\mathrm{s}}^{-1}]$ 
& $1.40^{+0.26}_{-0.21}$ &
$1.91^{+1.22}_{-0.34}$ & $1.33^{+0.36}_{-0.26}$ &
$0.94^{+0.18}_{-0.16}$\\ 
$E_{13\,{\mathrm{keV}}}$ [keV]&
$13.3^{+0.2}_{-0.2}$ & $13.6^{+0.5}_{-0.4} $ & $13.2^{+0.3}_{-0.2}$ &
$13.0^{+0.4}_{-0.7}$\\
 $\sigma_{13\,{\mathrm{keV}}}$ [keV] &
$2.8^{+0.4}_{-0.3}$& $4.0^{+1.0}_{-1.1}$& $3.4^{+0.4}_{-1.3}$ &
$3.2^{+0.8}_{-0.6}$ \\ 
$A_{13\,{\mathrm{keV}}} \dagger$ %[10^{-2}$ph ${\mathrm{cm}}^{-2}\,{\mathrm{s}}^{-1}]$ 
& $5.2^{+4.4}_{-1.6}$ &
$23.0^{+35.0}_{-14.0}$ & $7.0^{+5.3}_{-3.8}$ & $3.1^{+5.3}_{-1.3}$\\
$c_{\mathrm{HEXTE}}$ & $0.857^{+0.003}_{-0.003}$ &
$0.856^{+0.003}_{-0.003}$ & $0.854^{+0.003}_{-0.003}$ &
$0.860^{+0.004}_{-0.004}$\\ 
$L_{2-10\,{\mathrm{keV}}}\ddagger$ %[$10^{37}{\mathrm{erg\,s}}^{-1}$] 
& $5.4^{+0.4}_{-0.5}$ & $4.4^{+1.2}_{-1.1}$ &
$4.3^{+0.8}_{-0.3}$ & $4.0^{+0.6}_{-1.0}$\\ $\chi^2_{\mathrm{red}} /
{\mathrm{dof}}$ & $1.13 / 83$ & $0.71/ 83$ & $0.86/ 83$ & $0.77/ 83$
\\
\\ 
\multicolumn{2}{l}{$\dagger\,[10^{-2}$ph ${\mathrm{cm}}^{-2}\,{\mathrm{s}}^{-1}]$ }& 
\multicolumn{2}{l}{$\ddagger\,[10^{37}{\mathrm{erg\,s}}^{-1}$] }& 
\enddata 

\tablecomments{The data sets 1A--2B are defined in
Fig.~\ref{fig:light}. Parameters listed are the hydrogen column
density $N_{\mathrm{H}}$; the Fermi-Dirac cutoff and folding energies
$E_{\mathrm{cut}}$ and $E_{\mathrm{fold}}$; the CRSF energy
$E_{\mathrm{cyc}}$, width $\sigma_{\mathrm{cyc}}$, and depth
$\tau_{\mathrm{cyc}}$; the power law index $\Gamma$ and strength
$A_{\mathrm{pl}}$; the Fe K emission line energy $E_{\mathrm{Fe}}$,
width $\sigma_{\mathrm{Fe}}$, and strength $A_{\mathrm{Fe}}$; and the
13\,keV emission feature energy $E_{13\,{\mathrm{keV}}}$, width
$\sigma_{13\,{\mathrm{keV}}}$, and strength
$A_{13\,{\mathrm{keV}}}$. Uncertainties are quoted at the 90\%
confidence level assuming independent parameters (see text for
existing correlations). The flux normalization of the HEXTE instrument
with respect to the PCA, $c_{\mathrm{HEXTE}}$, and the luminosity in
the 2--10\,keV energy range derived from the PCA,
$L_{2-10\,{\mathrm{keV}}}$, are also given.}\label{tab:bestfit}
\end{deluxetable}

Based on modeling the averaged spectrum for the first part of orbit~1
(1A in Fig.~\ref{fig:light} and Table~\ref{tab:bestfit}) we find a
$2-10$\,keV unabsorbed flux of $7.0^{+0.5}_{-0.7} \times
10^{-9}$\,erg\,cm$^{-2}$\,s$^{-1}$ with the PCA.  Taking the deadtime
influence into account increases the measured flux by a few
percent. The \textsl{BeppoSAX} observation analyzed by B00 was
performed between 1997 February 27 19:45 and February 28 11:00 (UT)
and thus overlaps with parts of data set 1A of the \textsl{RXTE}
observation. B00 find a somewhat lower post eclipse egress 2--10\,keV
flux of $5.7\times10^{-9}$\,erg\,cm$^{-2}$\,s$^{-1}$.  The 18\%
discrepancy in flux is not surprising, due to the uncertainty in the
cross calibration between the PCA and the \textsl{BeppoSAX}. We point
out that the fluxes derived from the HEXTE instrument are 14\% lower
than those derived with the PCA (see $c_{\mathrm{HEXTE}}$ in
Table~\ref{tab:bestfit}) and thereby agree much better with
\textsl{BeppoSAX}.  Based on the PCA flux measurement we derived a
2--10\,keV luminosity of
$5.4^{+0.4}_{-0.5}\times10^{37}$\,${\mathrm{erg\,s}}^{-1}$, applying a
distance of 8\,kpc (B00:
$4.4\times10^{37}$\,${\mathrm{erg\,s}}^{-1}$). The fainter second
orbit shows a $2-10$\,keV flux of
$5.6^{+1.0}_{-0.4}\times10^{-9}$\,erg\,cm$^{-2}$\,s$^{-1}$ during its
first half, corresponding to a luminosity of
$4.3^{+0.8}_{-0.3}\times10^{37}$\,${\mathrm{erg\,s}}^{-1}$. The
luminosities after the mid-orbit drop are
$4.4^{+1.2}_{-1.1}\times10^{37}$\,${\mathrm{erg\,s}}^{-1}$ and
$4.0^{+0.6}_{-1.0}\times 10^{37}$\,${\mathrm{erg\,s}}^{-1}$ for orbit
1 and 2, respectively, i.e. the relative flux change in the second
orbit is a factor of two smaller than in the first orbit.

\section{Phase Averaged Spectroscopy}\label{sec:phase_avg}

\subsection{Time averaged spectral parameters}\label{sec:timeavg}

For each available ObsId ``standard2f'' mode PCA spectra and standard
science event mode HEXTE spectra were extracted, including the
corresponding background spectra. All spectra within one of the four
time intervals indicated in Fig.~\ref{fig:light}\emph{a} were
added. The resulting time averaged spectra are count rate selected in
the sense that they are characterized by episodes of different average
count rate levels within a given binary orbit, i.e., for each of the
two orbits the first interval spans the bright phase up to the
mid-orbit break, while the second interval spans the time up the
on-set of pre-eclipse dipping. Time averaged broad band source spectra
were modeled taking background subtracted PCA spectra from 3--23\,keV
and background subtracted HEXTE spectra from 17--100\,keV into
account, rebinning data above 60\,keV by a factor of five. As the CRSF
centroid energy is at $\sim 30$\,keV and a contamination of our data
due to the Xenon K-edge at 33\,keV would be possible, we decided not
to increase the PCA range up to 60\,keV as described by
\citet{rothschild06}, Appendix B.

\begin{figure}
\begin{center}
\plotone{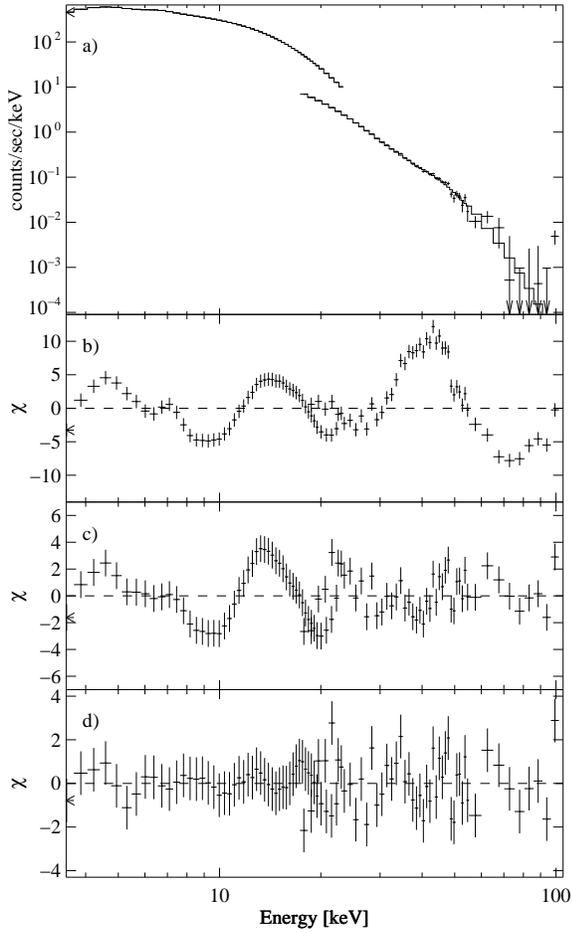}
\caption{(a) PCA and HEXTE phase averaged counts spectra, accumulated
  over the first part of the first binary orbit (1A in
  Fig.~\ref{fig:light} and Table~\ref{tab:bestfit}), and best fit
  model. (b) Residuals for an absorbed power law with a Fermi-Dirac
  cutoff, and a Gaussian Fe line. (c) Improved residuals obtained by
  adding a cyclotron line at $\sim$ 30\,keV. (d) Best fit residuals
  obtained by modeling the broad residual around 13\,keV with an
  additional Gaussian emission line. The reduced $\chi^2$ values are
  24, 3.4, and 0.97 from (b) to (d). See text for a detailed
  description of the model components and Table~\ref{tab:bestfit} for
  the best fit parameters.}
  \label{fig:spectrum}
\end{center}
\end{figure}

The basic continuum we use throughout the paper is an absorbed power
law with index $\Gamma$, modified by a Fermi-Dirac cutoff
characterized by cutoff and folding energies $E_{\mathrm{cut}}$ and
$E_{\mathrm{fold}}$ \citep[local XSPEC model \texttt{fdcut},][]{tanaka86}
\begin{equation}\label{eq:fdcut}
F(E) = A E^{-\Gamma}
\frac{1}{1+e^{(E-E_{\mathrm{cut}})/E_{\mathrm{fold}}}},
\end{equation}
a phenomenological model commonly applied to describe spectra of
accreting pulsars, e.g., \cite{kreykenbohm99}, \cite{kreykenbohm04},
and \cite{fritz06}. Fig.~\ref{fig:spectrum}\emph{b} shows the
residuals for fitting an absorbed \texttt{fdcut} model to the averaged
broad band spectrum for the first part of orbit 1 (1A in
Fig.~\ref{fig:light} and Table~\ref{tab:bestfit}), also including a
Gaussian line to model the narrow Fe K emission feature. The reduced
$\chi^2$ for this fit is $\chi^2_{\mathrm{red}}$=24 for 88 degrees of
freedom (dofs) and an absorption line is clearly visible in the
residuals at $\sim$ 25--30\,keV, which we interpret as a CRSF. To
model the CRSF we used a line shape with a Gaussian optical depth
profile which is the basis for past modeling of \textsl{RXTE} data
(local XSPEC model \texttt{gauabs}). The line is multiplicative with
respect to the continuum leading to the basic spectral model
\begin{equation}
f(E) = F(E) \times \exp{\Big(-\frac{\tau_{\mathrm{cyc}}}{2\pi
\sigma_{\mathrm{cyc}}}
e^{-\frac{1}{2}\big(\frac{E-E_{\mathrm{cyc}}}{\sigma_{\mathrm{cyc}}}\big)^2}\Big)},
\end{equation}
where $F(E)$ is given by Eq.~\ref{eq:fdcut}, and
$\tau_{\mathrm{cyc}}$, $E_{\mathrm{cyc}}$, and $\sigma_{\mathrm{cyc}}$
are the optical depth, centroid energy, and width of the cyclotron
line. While including this component improves $\chi^2_{\mathrm{red}}$
to 3.5 for 86 dofs a strong residual structure remains around $\sim$
13\,keV (Fig.~\ref{fig:spectrum}\emph{c}). Adding a single Gaussian
emission line at this energy results in a good overall fit with
$\chi^2_{\mathrm{red}} = 1.13$ for 83 dofs
(Fig.~\ref{fig:spectrum}\emph{d}).  The same model was also
successfully applied to describe the other three time-averaged spectra
with the resulting best fit parameters listed in
Table~\ref{tab:bestfit}.  We note that the line parameters do not
significantly depend on the continuum model chosen: fits using a
negative-positive exponential continuum \citep[local XSPEC model
\texttt{npex},][]{mihara95} instead of the power law with Fermi-Dirac
cutoff lead to the same cyclotron line parameters. Compared to a
\texttt{NPEX} or \texttt{highecut} continuum we achieved slightly
better spectral fits using a power law with Fermi-Dirac cutoff.
B00 used blackbody emission to describe the spectrum between
0.1 and 2\,keV. Our spectral data does not extend below 3.5\,keV and
we do not detect an influence of a black-body at lower
energies. \citet{becker07} are using a Comptonized Bremsstrahlung
spectrum to calculate the continuum. They state that the contribution
of blackbody emission is negligible in their spectrum.

To better understand the observed 13\,keV feature, we checked for
correlations for all possible pairs of spectral parameters using
$\chi^2$ contours.  We found no correlation between the parameters
characterizing the broad 13\,keV feature and the cyclotron line
parameters or the roll-over of the continuum model. As expected,
however, a dependence on the power law index $\Gamma$ (and therefore
also on $N_{\mathrm{H}}$) was clearly detectable.  An increasing
$\Gamma$ or $N_{\mathrm{H}}$ value resulted in a lower centroid energy
of the 13\,keV feature. We conclude that this feature is part of a
more complex continuum rather than the simple phenomenological
approach used here.

A broad feature at energies below the fundamental cyclotron line,
mostly around $\sim 10$\,keV, has also been observed in several other
cyclotron line sources, e.g., with \textsl{RXTE} in MXB~0656$-$072
\citep{mcbride06}, Her~X-1, 4\,U1626$-$67, 4U\,1907$+$09, and
4U\,1538$-$52 \citep{coburn02}, and with \textsl{Ginga} in
4\,U1538$-$52, 4U\,1907$-$09, and V\,0331$+$53 \citep{mihara95}.  A
weak 10\,keV absorption feature was also reported by \citet{santa98}
for a different \textsl{BeppoSAX} observation of Cen~X-3. Such
residuals are usually best described by an additional broad absorption
line to improve the fits. In some cases a broad emission line can
describe the data better. In our data we find that an emission line at
13\,keV describes the observed feature much better. Substituting the
13\,keV emission line by a Gaussian absorption line at $\sim$ 8\,keV
leads to $\chi^2_{\mathrm{red}}$ = 1.31 as compared to 1.13 for 83
dofs .

As described in \S\ref{sec:intro}, B00 observed quasi-simultaneously
with \textsl{BeppoSAX} (11 hours overlap with observation 1A from
Table~\ref{tab:bestfit}). A phase averaged analysis was carried out by
these authors using a power law with high-energy cutoff, smoothed
around the cutoff energy for the continuum. The CRSF parameters, with
$E_{\mathrm{cyc}}=30.6\pm 0.6$\,keV and $\sigma_{\mathrm{cyc}}=5.9\pm
0.7$\,keV, are comparable to those presented here .  Their measured
value of $N_{\mathrm{H}} = 1.95\pm0.03 \times
10^{22}\,{\mathrm{cm}}^{-2}$ is still within the error bars of our
RXTE observation of $N_{\mathrm{H}}
=1.6^{+0.4}_{-0.2}\,{\mathrm{cm}}^{-2}$. Their photon index of $\Gamma
= 1.208\pm0.007$ is much higher than our value of $\Gamma =
0.92^{+0.06}_{-0.04}$ due to our model definitions. We applied the B00
model to our data and obtained consistent results within error
bars. Due to a known deviation in the slope of the $\Gamma$-index
between \textsl{RXTE}-PCA and \textsl{BeppoSAX} \citep{kirsch05}, our
resulting value of $\Gamma = 1.26\pm 0.03$ for the B00 model is
marginally consistent.

\subsection{Physical model for CRSF}\label{sec:physmodel}

Ideally, models based on a physical description of the accretion
column above the magnetic poles of the neutron star should be used to
fit the shape of the cyclotron line and the spectral continuum.
Despite recent successes \citep{becker07}, however, a fully
self-consistent solution to the problem of continuum formation and
radiative transfer in neutron star accretion columns is not yet
available. Based on earlier work by \citet{araya99} and
\citet{araya00}, we have recently developed a model for the
self-consistent determination of cyclotron line profiles
\citep{schoenherr07} based on a prescribed continuum. Making use of a
Monte Carlo approach and using the correct quantum-electrodynamical
cross sections, \citet{schoenherr07} calculated the Green's function
for the radiative transport for several possible geometries of the
accretion column. By convolving an assumed input continuum spectrum
with these Green's functions, it is possible to calculate the photon
spectrum emerging from the accretion column. The magnetic field is
kept constant throughout the accretion column. This approach has been
implemented into XSPEC, allowing the direct comparison of the model
with observational data. We are presently working on an improved
version of the model with magnetic field gradients.

Here, we will present results from our analysis of the time averaged
20--70\,keV HEXTE data with this model. We limit ourselves to the
Green's functions in the 1-0 geometry for the line forming region
\citep{schoenherr07,freeman99,isenberg98b}, where the accretion column
is described as a slab with the slab normal being parallel to the
magnetic field and where the continuum spectrum is illuminating the
bottom of the slab.  This setup mimicks a plane-parallel and thin
emission region close to the surface. The continuum is again taken to
be a power law modified by a Fermi-Dirac cutoff (Eq.~\ref{eq:fdcut}).
We decided to excluded $N_{\mathrm{H}}$ from the continuum because it
is neglectable in the chosen energy range A gravitational redshift
$z=0.3$ is assumed, a typical value for neutron stars (see
\S\ref{sec:intro}).  A detailed analysis of several different sources,
including phase resolved analysis, will be presented in a future
paper.

\begin{figure}
\plotone{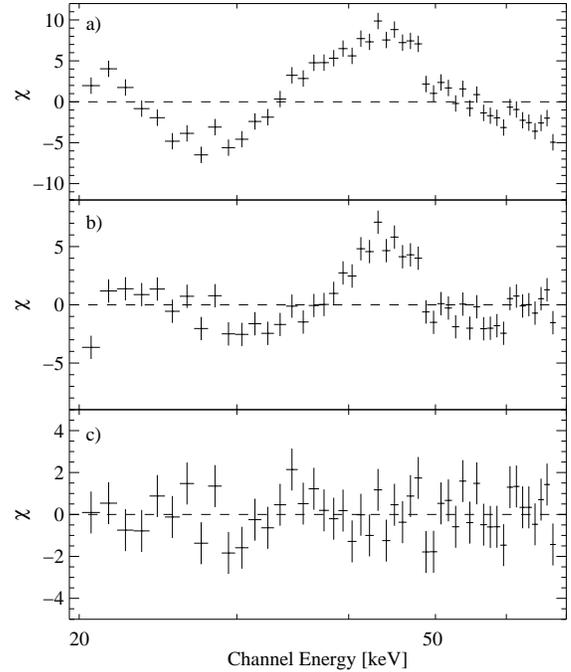}
\caption{Residuals for fitting data with CRSF shapes obtained for
  Monte Carlo simulations applied on phase averaged HEXTE data from
  20-80\,keV. (a)~continuum \texttt{fdcut*powerlaw} with
  $\chi^2_{\mathrm{red}}=24.0$ (45 dofs). (b)~included
  \texttt{cyclomc} model for Slab 1-0 geometry showing emission wings
  ($\chi^2_{\mathrm{red}}=8.0$ with 41 dofs). (c)~added partial
  covering to reduce emission wings ($\chi^2_{\mathrm{red}}=1.4$ with
  39 dofs). Note that the residuals in (a) are slightly
  different than those shown in Fig.~\ref{fig:spectrum}\,a since only
  the HEXTE data were modeled.}
\label{gabiplot}
\end{figure}

Fig.~\ref{gabiplot}a shows the residuals after modeling the HEXTE
spectrum with the continuum model without taking the cyclotron line
formation into account. Introducing the cyclotron line model and
refitting significantly reduces the residuals (Fig.~\ref{gabiplot}b),
although some structure remains. As has been shown by
\citet{schoenherr07}, this structure is caused by the fact that the
predicted self-consistent cyclotron line shapes are generally too deep
and in addition have significant emission wings, assuming the
simplified scenario of a constant magnetic field and simple
geometry. These wings are due to so-called spawned photons, which are
the result of the stepwise de-excitation of electrons from higher
Landau levels after their excitation through a scattering photon
\citep{araya99,araya00}. One way of reducing the line depth and
strength of the wings is by invoking a partial covering approach,
where only part of the seed photon spectrum is seen through the
accretion column, while the remainder is observed without being
modified by the column.  Note that this approach is not unique. Other
geometries, such as a seed photon source distributed throughout the
accretion column, a column with a temperature gradient, or a
vertically varying magnetic field, could also result in shallower
cyclotron line profiles with less dominant wings \citep{schoenherr07}.

The result of fitting the data with a partial covering model is shown
in Fig.~\ref{gabiplot}c. This model describes the data well
($\chi^2_{\mathrm{red}} = 1.35$ for 40 degrees of freedom). The best
fit parameters are $B=3.46^{+0.07}_{-0.03}\times 10^{12}$\,G for the
magnetic field, $k T_{\mathrm{e}}=7.04^{+0.63}_{-0.48}$\,keV for the
electron temperature, $\tau_{\mathrm{T}} = 2.4^{+0.05}_{-0.07}\times
10^{-3}$ for the Thomson optical depth of the continuum, not to be
confused with the optical depth of the \texttt{gauabs} model. Note
that the optical depth in the lines can be larger by a factor of more
than $10^4$ \citep{araya99}. The cosine of the angle, $\theta$,
between our line of sight and the magnetic field is
${\mathrm{cos}}(\theta)=0.94^{+0.00}_{-0.07}$, i.e.,
$\theta=20^{+10}_{-0}$\,${}$ degrees. 

To check the consistency of our result with that determined in
\S\ref{sec:timeavg}, we use the best-fit parameters from above to
calculate the centroid energy of the cyclotron line
\citep[][Eq.~3]{schoenherr07}
\begin{equation}
E_{\mathrm{cyc}} = m_{\mathrm{e}}c^{2}\frac{\sqrt{1+2B/B_{\mathrm{crit}}\,
{\mathrm{sin}}^2(\theta)}-1}{{\mathrm{sin}}^2(\theta)} \times \frac{1}{1+z}
\end{equation}
giving $E_{\mathrm{cyc}}=30.9^{+0.7}_{-0.3}$\,keV for data set~1A. The
width of the CRSF can be estimated from Doppler broadening
\citep[e.g.,][]{meszaros85}
\begin{equation}
\Gamma_{\mathrm{FWHM}} = \sqrt{\frac{8 {\mathrm{ln}}(2)
kT_{\mathrm{e}}}{m_{\mathrm{e}}c^2}}\left|{\mathrm{cos}}(\theta)\right|E_{\mathrm{cyc}}.
\end{equation}
In our case we obtain a width of $8.0^{+0.5}_{-0.9}$\,keV. Both
results lie within the uncertainties of Table~\ref{tab:bestfit}.  The
CRSF depth in this model is described by the Thomson optical depth and
cannot be directly translated into the \texttt{gauabs} depth due to
strong dependencies of the scattering cross section on the emission
angle $\theta$.

\subsection{Spectral evolution over the orbit} \label{sec:evolution}

This data set allows us to perform the most complete and detailed
analysis of the orbital dependence of Cen~X-3's broad band spectrum to
date. We extracted PCA and HEXTE spectra from the same data modes as
for the time averaged analysis (\S\ref{sec:timeavg}) but using shorter
GTIs with their duration mainly constrained by observation gaps due to
SAA passages and earth occultations. This selection procedure leads to
spectra with a typical exposure time of $\sim$ 3.5\,ks ($\Delta
\phi_{\mathrm{orb}} \sim 0.03$). Data subsets with considerably longer
observation times were further split into subsets of $\sim$ 3.5\,ks
duration. We ignored data taken during the eclipse and we also ignored
the data set at the end of the pre-eclipse dip in orbit~1, since no
HEXTE data were available.  The resulting broad band spectra were then
modeled with the same model used to describe the time averaged spectra
in \S\ref{sec:timeavg}.  The chosen energy range was 3.5--23\,keV for
PCA and 18--60\,keV for HEXTE.

\begin{figure}
\plotone{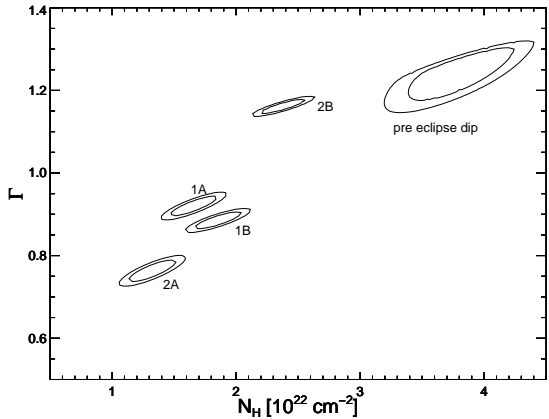}
\caption{$\chi^2$ contours for the four phase averaged spectra
presented in Table~\ref{tab:bestfit} and the pre-eclipse dip. For
each data set the 1- and 2-$\sigma$ confidence levels are shown. Note
that the individual $N_\mathrm{H}$--$\Gamma$ correlation is a
non-physical effect described in the text.}
\label{fig:contours}
\end{figure}

Fig.~\ref{fig:light}\emph{b}--\emph{e} shows selected time-resolved
spectral parameters. As already indicated by the time averaged results
(Table~\ref{tab:bestfit}), the two orbits show a somewhat different
behavior with respect to the two luminosity levels observed in each
case.  During the first orbit the parameters are essentially stable
within the uncertainties over most of the orbit.  $N_{{\mathrm{H}}}$
increases near eclipse ingress and egress during both orbits
(Fig.~\ref{fig:light}\emph{b}) as expected from an increasing amount
of stellar material in the line of sight \citep{clark88}. During the
pre-eclipse dip in orbit~1, $N_{{\mathrm{H}}}$ is enhanced by a factor
of $\sim$ 2.  $\Gamma$ also increases during the pre-eclipse dip
(Fig.~\ref{fig:light}\emph{c}).  Error contours for the four phase
averaged observations from \S\ref{sec:timeavg} and for the pre-eclipse
dip in the first orbit are shown in Fig.~\ref{fig:contours}.  The
observed $N_{\textrm{H}}$--$ \Gamma$ correlation for the individual
data sets is real and can not be sufficiently explained by the
calculated error contours. This correlation is of unknown origin. The
$N_{\textrm{H}}$--$\Gamma$ correlation observed for each individual data set is
an artificial effect from the continuum modeling. When modeling the
spectrum one achieves similar $\chi^{2}$ values for different
$N_{\textrm{H}}$--$ \Gamma$ combinations, since increasing
$N_{\textrm{H}}$ can compensate for a softer power law component.
Therefore a non-physical correlation in $N_{\textrm{H}}$ and $\Gamma$
is expected.

During the second orbit we observe a change of the general behavior of
many of the spectral parameters that is apparently related to the
mid-orbit luminosity drop. $N_{{\mathrm{H}}}$ stays relative constant
prior to mid-orbit and rises continuously afterward
(Fig.~\ref{fig:light}\emph{b}). Analysis with a frozen $\Gamma$ value
to take the $N_{{\mathrm{H}}}$-$\Gamma$-correlation into account,
showed the same tendency in $N_{{\mathrm{H}}}$ during the second
orbit. As in the pre-eclipse dip of orbit~1, the power law $\Gamma$
slightly increases with $N_{{\mathrm{H}}}$
(Fig.~\ref{fig:light}\emph{c}). The CRSF parameters variability is
slightly more pronounced during the second orbit
(Fig.~\ref{fig:light}\emph{d} and Fig.~\ref{fig:light}\emph{e}). Note
that the cutoff and folding energies do not show any significant
change throughout the whole observation.

\subsection{Wind model}\label{sec:windmodel}

\begin{figure}
\plotone{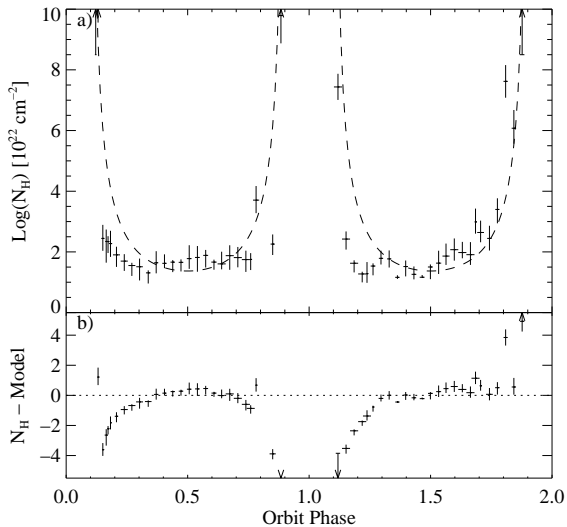}
\caption{(a) values of $N_{{\mathrm{H}}}$ as a function of orbital phase
(crosses) and calculated model for parameters mentioned in text
(dashed line). (b) Difference of $N_{{\mathrm{H}}}$ and wind model.}
\label{windplot}
\end{figure}

We calculated a simple radiation driven wind model for
$N_{{\mathrm{H}}}$ based on the model by \citet[][hereafter
CAK]{castor75}, although we note that this assumption of a radiation
driven wind is only an approximation, as it is very likely that the
wind is disrupted by the strong X-rays present in the system
\citep[and references therein]{wojdowski01, wojdowski03}. We assumed a
mass loss, $\dot{m}$, for the donor star and the CAK velocity profile
$v(r)=v_{\infty}(1-R_\star/r)^\beta$, where $v_{\infty}$ is the
terminal velocity. This gives
\begin{equation}
  n_{{\mathrm{H}}}(r)=\frac{\dot{m}}{4\pi k r^2 v(r)}
\end{equation}
for the radial particle density profile, where $k$ is a conversion
factor from mass to effective proton particle density, assuming a wind
with cosmic abundances, a system inclination of $90^\circ$, and
ignoring the orbital eccentricity ($e\leq 0.016$).  We then integrate
over the density along the line of sight for a specific position in
the orbit:
\begin{equation}
  N_{\mathrm{H}} = \int_{l_{\mathrm{NS}}}^{\infty}n_{{\mathrm{H}}}(l')dl',
\end{equation}
where $l_{{\mathrm{NS}}}= R_{\mathrm{d}}
{\mathrm{cos}}(\phi_{\mathrm{orb}})$ is the position of the neutron
star, projected to the celestial plane and where $R_{\mathrm{d}} =
19\,R_\odot = 1.58\,R_\star$ is the separation between neutron star
and donor \citep{blondin94}. Assuming a typical value of $\beta =0.8$
\citep{friend86} we obtain good results for $\dot{m}=
7\times10^{-7}\,\mathrm{M}_\odot\,{\mathrm{yr}}^{-1}$ and $v_\infty =
1200\,{\mathrm{km\,s}}^{-1}$ (Fig.~\ref{windplot}\emph{a}). For better
comparison we also show the difference between data and model
(Fig.~\ref{windplot}\emph{b}). Due to the definition of the model a
strong correlation between increasing mass loss and terminal velocity
is present, such that absolute values cannot be determined by this
approach.

Both orbits show a different behavior in the evolution of
$N_{{\mathrm{H}}}$. During the first orbit, where $N_{{\mathrm{H}}}$
is mostly constant, we see a symmetrical deviation during eclipse
ingress and egress, when our simple approach fails to model the
data. A similar deviation is seen at the beginning of the second
orbit. After mid-orbit $N_{\mathrm{H}}$ keeps increasing for the
remainder of the orbit, indicating an increased absorption in the
second half of the orbit. A similar behavior has been simulated by
\cite{blondin90} and probably indicates the presence of an accretion
wake in this system. A small bump around $\phi_{\mathrm{orb}}
=0.3-0.4$ can be observed in both orbits, more pronounced in the
second one. This bump also exists in the simulation and is another
indication of a tidal wake.

\section{Phase resolved spectroscopy}\label{sec:phase_res}
\subsection{Pulse profile} \label{sec:pulseprofile}

\begin{figure}
\plotone{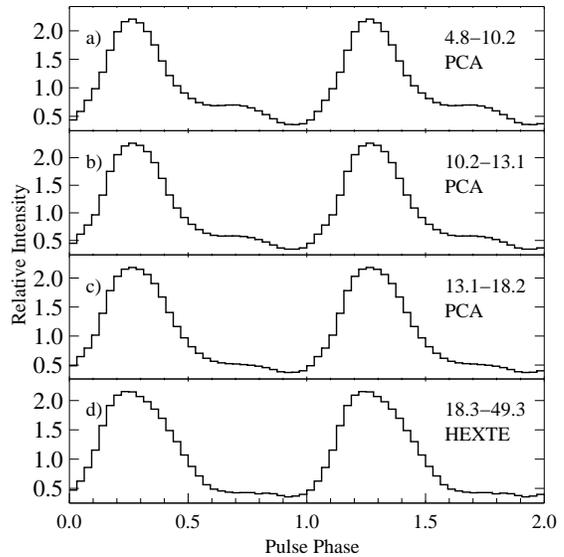}
\caption{Normalized PCA and HEXTE pulse profiles for four energy
bands: (a)~4.8--10.2\,keV, (b)~10.2--13.1\,keV, (c)~13.1--18.2\,keV
and (d) 18.3--49.3\,keV. A secondary peak at pulse phase $\sim 0.7$ is
visible for lower energies.}
\label{fig:profile}
\end{figure}

Pulse profiles with 32 phase bins for 4 energy bands were extracted
for each of the 12 ObsIds. We decided to use locally determined pulse
periods for each ObsID and the ephemeris from \citet{nagase92}, after
using the ephemeris and period from B00 did not result in consistent
pulse profiles. All pulse profiles showed a similar shape with a main
peak at pulse phase ($\phi_{\mathrm{pulse}}$) $\sim 0.3$, which
allowed us to combine them after phase alignment.
Figure~\ref{fig:profile} shows the resulting pulse profiles for four
different energy ranges.  A second weak peak at pulse phase of $\sim
0.7$ is visible at energies below 18\,keV.  \citet{nagase92} observed
Cen X-3 in 1989 March with \textsl{Ginga} when the source had a
$1-37$\,keV luminosity of $5 \times
10^{37}$\,${\mathrm{erg\,s}}^{-1}$, a factor of two fainter than
during this RXTE observation, and presents pulse profiles for 9 energy
bands.  At lower energies they observe a double peaked profile, where
both peaks have equal intensities below 7\,keV. Above 18\,keV the
profile is comparable to our results with only one main
peak. \citet{audley01} showed that the $2-25$\,keV pulse profile is
highly variable, changing between an asymmetric double peaked, as in
our observation, and a much more complex pulse profile with multiple
peaks, depending on the count rate of the binary system.

\begin{figure}
\plotone{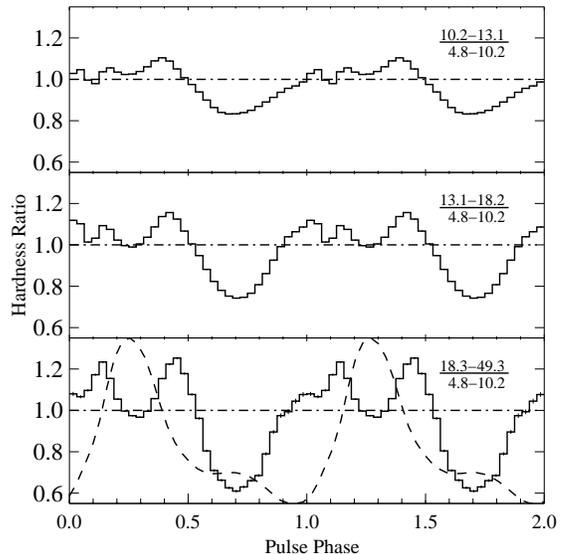}
\caption{Normalized hardness ratios. Each panel has been calculated by
  taking the according pulse profile from Fig.~\ref{fig:profile}
  divided by panel (a) from Fig.~\ref{fig:profile}. The pulse profile
  is indicated in the bottom panel (dashed line). Top right corner
  show the energy ranges.}
\label{fig:hardness}
\end{figure}

We used the pulse profiles presented in Fig.~\ref{fig:profile} to
create hardness ratios for each energy band using the lowest energy
range as baseline (Fig.~\ref{fig:hardness}). At $\phi_{\mathrm{pulse}}=
0.7$ we see the decrease in hardness of the secondary peak toward
higher energies. The rise and decay of the main peak
($\phi_{\mathrm{pulse}}=0.1$ and $0.5$) show peaks in the hardness ratio,
a clear indication that the lower energy emission region is beamed
more sharply. It is not clear if a third peak, visible at
$\phi_{\mathrm{pulse}}= 1.0$, originates from a real physical feature or
is part of an artificial wing from a phase continuity in the plot.

\subsection{Spectral evolution over the pulse profile}\label{sec:pulsevolution}

\begin{figure}
\begin{center}
\plotone{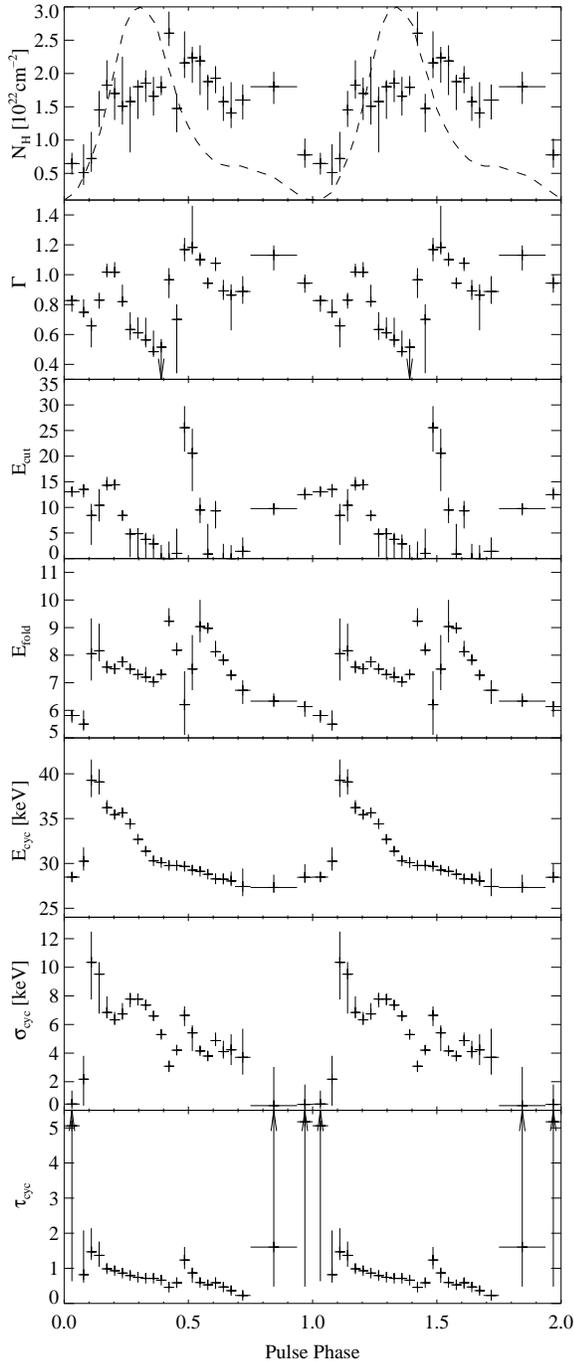}
\caption{Selected spectral parameters over the pulse profile (dotted
  line). The last phase bins are binned. Error bars indicate 1$\sigma$
  deviations. }
\label{fig:resolvedspec}
\end{center}
\end{figure}

Extraction of phase resolved spectra for each individual ObsId with 32
phase bins resulted in large uncertainties for the individual
parameters.  We therefore created phase resolved spectra for each
ObsIDs with 8 phase bins and observed similarities in the behavior of
the spectral parameters, e.g., an increase of the CRSF centroid energy
during the rise of the main peak. We decided to sum all ObsIDs and use
32 phase bins resulting in an integration time of $\sim$12\,ks for a
phase analysis. In the pulse minimum we binned the phases by a factor
of 2, and even by a factor of 6 for the secondary peak, to be able to
constrain the CRSF feature. This results in a total of 24 phase bins
on which we performed a spectral analysis using the model defined in
\S\ref{sec:timeavg}.  Fig.~\ref{fig:resolvedspec} shows the best-fit
values of selected parameters over the pulse period. Compared to the
phase averaged analysis, not only N$_{{\mathrm{H}}}$ and $\Gamma$ but
also the CRSF parameters change drastically over the pulse period.

The pulse profile can be roughly divided into three regions where the
continuum parameters are significantly different: main peak, secondary
peak, and pulse minimum.  The column density $N_{\mathrm{H}}$ is
constant with a value of
$1.5_{-0.4}^{+0.3}\times10^{22}\,{\mathrm{cm}}^{-2}$ during the main
peak. Between main and secondary peak, this value suddenly jumps up,
to $2.2_{-0.3}^{+0.2}\times10^{22}\,{\mathrm{cm}}^{-2}$.  $\chi^2$
contours indicate that this increase is not due to statistical
fluctuations. During the pulse minimum the $N_{\mathrm{H}}$ value
reaches a minimum of
$0.5_{-0.2}^{+0.5}\times10^{22}\,{\mathrm{cm}}^{-2}$.  Another
interpretation would be that $N_{\mathrm{H}}$ follows the shape of the
pulse profile and is somehow capped during the main peak.  We have not
identified a physical explanation for this observational fact which
might be caused by the limitations of the current continuum modeling.
The power law index $\Gamma$ also changes strongly over the pulse
phase.  It increases during the rise of the pulse up to a value of
$1.0_{-0.1}^{+0.1}$ and then decreases to $0.5_{-0.3}^{+0.2}$ for the
duration of the main peak. After the main peak the value jumps back up
to $1.2_{-0.1}^{+0.1}$.  The CRSF centroid energy also shows a drastic
change throughout the pulse.  During the rise of the main peak the
value increases from $\sim30$\,keV to almost 40\,keV, slowly
decreasing throughout the main peak and descent below the phase
average value of less than 30\,keV. The evolution of the CRSF
throughout the pulse is similar to the shape of the pulse profile with
a phase shift of $\Delta\phi_{\mathrm{pulse}}=0.2$ (ignoring the two
outliers in the pulse rise we still obtain
$\Delta\phi_{\mathrm{pulse}}=0.1$). The width $\sigma_{\mathrm{cyc}}$
of the CRSF approximately follows the evolution of the centroid energy
over the main peak, and drops to $\sim 0$ during the off-pulse, an
indication that the CRSF is not well constrained. The depth
$\tau_{\mathrm{cyc}}$ is slowly decreasing throughout the whole pulse
and is also much less constrained during the pulse minimum.  Compared
to the other known sources featuring a variation of the CRSF, e.g.,
Her X-1 \citep{gruber01}, Vela~X-1 \citep{labarbera03,kreykenbohm02},
4U\,0352$+$309 \citep{coburn01}, and \mbox{GX~301$-$2}
\citep{kreykenbohm04}, Cen X-3 is the only known source to show an
increase of the centroid energy during the ascent of the main pulse.

B00 also provided a phase resolved analysis, but they divided the
pulse into only 4 parts: ascent, maximum, descent of the main peak,
and minimum, including the secondary peak. They also observe an
increase in the CRSF centroid energy up to $36.6^{+1.6}_{-2.4}$\,keV
in the ascent.  We combined the phase resolved spectra according to
the phase bins in B00 and applied their model on our results. We can
confirm the results from B00 within the error bars with small
deviations for $\Gamma$.

\section{Discussion and Conclusions} \label{sec:discussion} In this
paper we presented the data analysis of the HMXB Cen X-3 observed for
two consecutive orbits. We conducted phase averaged analysis for
different parts of the orbit, distinguished by a drop in the overall
count rate. A decrease in the 2--10\,keV luminosity between first and
second orbit by $\sim 20\%$ can be observed and could be caused by a
decrease in the overall amount of matter accreted onto the neutron
star. In addition, pulse phase resolved spectroscopy was
performed. Here, the long duration of the observations allowed us to
study the pulse phase dependence of the cyclotron line parameters in
unprecedented detail.  In the following two sections, we summarize the
major results of these two studies, starting with the orbital
variability of the X-ray spectrum.

\subsection{Orbital Variability of the X-ray Spectrum }

The analysis for individual satellite orbits shows the evolution of
the spectral parameters throughout both orbits. We see a difference
between both orbits, not only in the luminosity, but also in the
evolution of $N_{\mathrm{H}}$ and $\Gamma$, both correlated to each
other. Comparison with a simple CAK wind model showed that, while
reproducing the overall trends in $N_{\mathrm{H}}$, such a model is
not sufficient to describe the observed data fully, regardless of
inaccuracies during eclipse ingress and egress. \citet{wojdowski01}
use a similar approach on a Cen~X-3 data set from \textsl{ASCA} with
significantly different results. Probing the parameter space they find
best fit values for $\beta=0.57^{+0.06}_{-0.07}$ and
\begin{equation}
\Xi=\frac{\dot{M}}{10^{-6}\,{\mathrm{M}}_\odot\,{\mathrm{yr}}^{-1}}
\left(\frac{v_\infty}{1000\,{\mathrm{km}}\,{\mathrm{s}}^{1}}\right)^{-1}
\left(\frac{d}{10\,{\mathrm{kpc}}}\right)^{-1}
\end{equation} 
with $\Xi = 1.56$. Applying these parameters to our model we find a
minimum $N_{\mathrm{H}}$ value of $3.2 \times
10^{22}\,{\mathrm{cm}}^{-2}$, 1.5 times higher than our observed
data. Also the change of $N_{\mathrm{H}}$ during ingress and egress of
the eclipse is not as drastic as observed here.  With a fixed $\beta =
0.8$ and $\Xi= 0.7$ we obtain the results shown in
Fig.~\ref{windplot}.

In our data set a small increase in $N_{\mathrm{H}}$ is seen in the
first half ($\phi_{\mathrm{orb}}=0.3-0.4$) in both orbits, more
pronounced in the second orbit. \cite{blondin91} used two-dimensional
numerical simulations based on the CAK model to calculate the column
density in massive X-ray binary systems with a tidal stream and
observes such an increase in $N_{\mathrm{H}}$ around
$\phi_{\mathrm{orb}}=0.4-0.5$. They interpret this as a leading bow
shock in front of the accretion stream which is extremely sensitive to
orbital parameters, e.g., distance between neutron star and donor or
the intensity of the stellar wind. A possible explanation of our
observed increase is that it corresponds to this bow shock observed in
the simulations. In the second half of the orbit, \cite{blondin91} see
a general increase in $N_{\mathrm{H}}$, comparable to our observations
during the second orbit. This increase could be due to material in the
line of sight, trailing the tidal wake. Taking into account that the
continuum parameters do not change significantly over the orbit we can
rule out that the system changed its state.  Similar increases have
been observed in other sources, e.g., 4U1700$-$37 \citep{haberl89} and
Vela X-1 \citep{haberl90}, both systems are believed to have a tidal
streams. In these cases, the difference was more than an order of
magnitude, whereas here the increase is on the order of a factor of
2--3. We point out that we also observe a change in $N_{\mathrm{H}}$
with a fixed photon index so that the
$N_{\mathrm{H}}$-$\Gamma$-correlation alone can not be responsible for
this rise. The fact that this increase is only visible during the
second orbit could imply that the overall luminosity has an influence
on the visibility of the accretion wake.

We do not rule out other possible explanations for the differences in
orbit 1 and 2.  \citet{blondin90} showed in their simulations that the
column density throughout the orbit changes between consecutive orbits
due to variability in the accretion flow. The drop in luminosity of
$\sim20$\,\% between orbit 1 and 2 indicates the decrease of
$\dot{m}$. In the simulations \cite{blondin91} observe instabilities
for lower luminosities, i.e., less photon pressure from the donor
star, trailing the tidal wake. These instabilities can create small
pockets of denser material which also could be responsible for the
observed increase in $N_{\mathrm{H}}$.

Another hint for the existence of multiple material clumps is seen in
the light curve. Compared to the first orbit, the second orbit shows
multiple pre-eclipse dips. In the first orbit, during the pre-eclipse
dip $N_{{\mathrm{H}}}$ increases by a factor of $\sim 2$. Multiple
dips close together could cause an overall increase of
$N_{\mathrm{H}}$ throughout the second orbit. One small drawback to
this theory is that the increase in $N_{\mathrm{H}}$ starts directly
after the mid-orbit break, whereas the pre-eclipse dips are occurring
at $\phi_{\mathrm{orb}}>0.7$ of the second orbit. Therefore we do not
think that pre-eclipse dips can be exclusively responsible for the
rise.

\subsection{Spectral and Pulse Phase Variability}

We divided each orbit into two parts, before and after mid-orbit
break, and extracted phase averaged spectra for each part. For the
analysis we applied a power law with a Fermi-Dirac cutoff as continuum
with an additional iron emission line. A CRSF at $\sim 30$\,keV is
visible in all data sets, consistent with previous observations. The
widely used local model \texttt{gauabs} gave us best fit results for
this feature. No significant changes in the spectral parameters could
be observed throughout the orbit. An additional broad Gaussian
emission feature at 13\,keV had to be introduced to get reasonable
residuals. We checked for correlations of this feature with other
model parameters and determined that it is probably an artifact of the
current continuum modeling. We conclude that the 13\,keV feature is
equal to the 10\,keV absorption feature that has been observed in
multiple different sources and with different instruments.  Further
investigation of this feature is necessary, but this is beyond the
scope of this paper.

We also applied a newly developed physical model, \texttt{cyclomc}
\citep{schoenherr07}, based on Monte-Carlo simulations to part of the
phase averaged data (data set 1A). This model is self-consistent in
determining the CRSF parameters using the Green's function for
radiative transport. For all geometries of the line forming region
available in the current implementation of the model, we generally
find that the predicted CRSFs are too deep and possess significant
emission wings, leading to difficulties when matching the observed
spectra. The slab 1-0 geometry, where the continuum is illuminating
the bottom of the accretion column, has the smallest emission wings of
the geometries tested and is therefore a good candidate for a first
approach. Assuming partial covering of the continuum leads to a
further reduction of the remaining emission wings. This approach is
motivated by the idea that only a part of the continuum contribution
emerges from the line forming region.  The results are encouraging and
in good agreement with the previous phenomenological approach. It is
planned to improve this model by, e.g., adding magnetic field that are
gradually changing over the accretion column, and to apply it to
different sources with known CRSFs as part of a future paper.

For the phase resolved analysis we created pulse profiles in different
energy bands over the whole observation. They show a main peak and a
much smaller secondary peak which is only seen at lower energies.
Compared to other sources, e.g., Vela X-1 or GX 301$-$2, the pulse
profile is rather smooth and does not show a complex shape at lower
energies. The main peak is more pronounced at lower energies and
slightly asymmetric, steeper in the rise, at higher energies.
Comparing with \citet{nagase92}, where a clear double peaked profile
at lower energies has been observed, we assume that this is a
luminosity dependent effect on the pulse profile.  During the
observation of \citet{nagase92}, Cen X-3 was fainter than during our
second orbit and no changes in the pulse profiles have been observed
throughout our observation.

For each bin in the pulse profile we extracted a spectrum to study the
evolution of the parameters over the pulse. For better statistics we
combined some of the data in the pulse minimum, resulting in 24 phase
bins overall. The continuum parameters show some significant changes
throughout the pulse and the general $N_{\mathrm{H}}$-$\Gamma$-correlation
seems not to be valid during the main peak.

\begin{figure}
\plotone{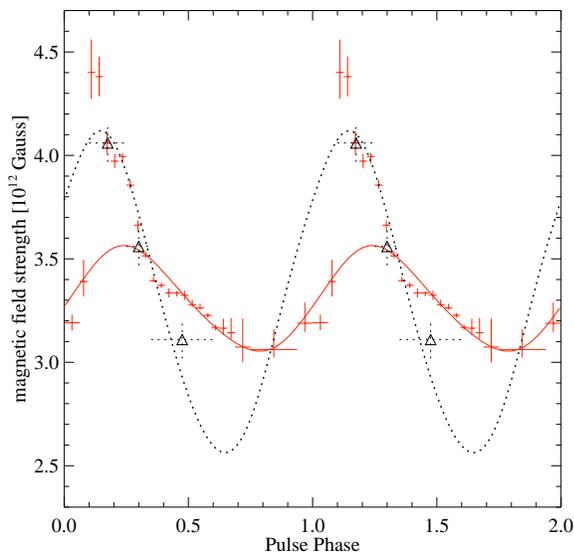}
\caption{Calculated magnetic field strengths (crosses) and B00 results
(triangles) using the CRSF centroid energy. The dotted line
show the single dipole model using parameters from B00. The solid
line show the best fit of the single dipol model to our data.}
\label{fig:dipol}
\end{figure}

The CRSF shows significant changes throughout the pulse. The shape of
the changes is very similar to the pulse profile shifted by
$\Delta\phi_{\mathrm{pulse}}=0.1$, not including the two maximum
values during the rise. These outliers have rather large uncertainties
and should be handled with caution in the interpretation. From the
main peak on, $E_{\mathrm{cyc}}$ slowly decreases to below 30\,keV
where it stays for the remainder of the pulse. Fig.~\ref{fig:dipol}
shows how the magnetic field deduced from these measurements varies as
a function of pulse phase. The figure also shows the earlier
\textsl{BeppoSAX} measurements of B00, which are consistent with our
much finer resolved data.

B00 argued that the large variation in $B$-field with pulse phase
could not be due to a variation of the height in the accretion column
at which the observed radiation originates, since this variation would
imply height differences of kilometers. B00 instead argue that, based
on earlier studies of the pulse shape of Cen~X-3 by \citet{leahy91},
the polar cap of Cen~X-3 is rather large. If this assumption is true,
then it is likely that different locations on the neutron star's
surface are observed over one rotation of the neutron star.  Using the
assumption that the observed X-rays are only produced by one of the
two magnetic poles, and furthermore assuming that the observed X-rays
are all coming from a location at a constant geographical colatitude
of the neutron star (taken by B00 to be the system's inclination), B00
then show that the variation of the $B$-field at the neutron star's
surface produced by a magnetic dipole offset by $0.1\,R_{\mathrm{NS}}$
from the rotational axis of the neutron star can explain the $B$-field
variation seen by \textsl{BeppoSAX}. We reproduce their best-fit model
in Fig.~\ref{fig:dipol}. While the model is sufficient to explain the
\textsl{BeppoSAX}-data, the higher resolution \textsl{RXTE}-data
clearly show that it is not a viable explanation for the observed
$B$-field variation. We extended the approach of B00 by allowing for
observations of both magnetic poles and by also removing the
constraint that the colatitude at which the X-rays originate equals
the inclination. Neither of these approaches resulted in statistically
satisfying descriptions of the variation of $B$ with
$\phi_{\mathrm{pulse}}$. Even when relaxing the model assumptions
further, by using a skew-symmetric magnetic dipole, which approximates
higher multipole moments for the $B$-field on the surface, and
assuming that the X-rays come from two different locations on the
neutron star, no satisfying description of the observed line variation
was found. We therefore conclude that the model of B00 is not a viable
description of the high resolution $B$-field variation seen here.

We note that an alternative study of the pulse profile of Cen~X-3 has
been presented by \citet{kraus96}. Similar to the assumptions outlined
above, these authors show that the shape and energy dependence of the
pulse profile can be explained by emission from a distorted magnetic
dipole. Taking into account the relativistic photon propagation close
to the neutron star, \citet{kraus96} show that, in order to explain
the pulse profile, both magnetic poles must contribute to the observed
emission. Modeling observations of Cen~X-3 made at a similar
luminosity state as the one during the \textsl{RXTE} observation,
\citet{kraus96} show that these assumptions yield an offset of
approximately $10^\circ$ for both poles, and that both poles
contribute equally to the observed flux. In these models, they assume
that the polar caps radiation characteristic is a pen- plus
pencil-beam pattern. A major result of these models is that in this
decomposition the main pulse emission is dominated by the magnetic
pole facing toward the observer, while the wings of the main pulse are
dominated by the pole facing away from the observer. Although they do
not take the height of the accretion column into account, we believe
it is likely that for geometric reasons the X-rays observed from the
magnetic pole on the neutron star hemisphere facing away from the
observer originate at a distance farther away from the neutron star
surface, and thus in a region of lower $B$-field. Together with the
possibility of slightly different surface magnetic strengths at both
poles, we therefore deem it likely that the observed $B$-field
variation over the X-ray pulse is due to us observing the two
accretion columns at different heights, although further modeling of
such a geometric setup is clearly required before a final answer on
the origin of the $B$-field variation can be given.

\begin{acknowledgments}
  We acknowledge the support of NASA contract NAS5-30720, DLR contract
  50OR0302, a travel grant from the Deutscher Akademischer
  Austauschdienst, and support provided by the Studienstiftung des
  Deutschen Volkes to VG and GS. We thank the members of the pulsar
  team supported by the International Space Science Institute (ISSI)
  in Bern, Switzerland, for discussions which greatly helped shape the
  ideas presented in this paper, and ISSI itself for its hospitality.
\end{acknowledgments}

\end{document}